\documentclass[lettersize,journal]{IEEEtran}
\usepackage{amsmath,amsfonts}
\usepackage{algorithmic}
\usepackage{algorithm}
\usepackage{array}
\usepackage[caption=false,font=normalsize,labelfont=sf,textfont=sf]{subfig}
\usepackage{textcomp}
\usepackage{stfloats}
\usepackage{url}
\usepackage{verbatim}
\usepackage{graphicx}
\usepackage{cite}
\usepackage{multirow}
\begin{document}
\title{Understanding Memristive Behavior: An Atomistic Study of the Influence of Grain Boundaries on Surface and Out-of-Plane Diffusion of Metallic Atoms}%

\author{Mohit D. Ganeriwala, Daniel Luque-Jarava, Francisco Pasadas, Juan J. Palacios, Francisco G. Ruiz, Andres Godoy and Enrique G. Marin%
        % <-this % stops a space
\thanks{Mohit D. Ganeriwala, Daniel Luque-Jarava, Francisco Pasadas, Francisco G. Ruiz, Andres Godoy and Enrique G. Marin are affiliated with the Department of Electronics and Computer Technology, Universidad de Granada, Granada, Spain.(e-mail: mohit@go.ugr.es, egmarin@go.ugr.es)}%
\thanks{Juan J. Palacios is with Departamento de Física de la Materia Condensada, Universidad Autónoma de Madrid, 28049 Madrid, Spain.}}%


%Mohit D. Ganeriwala, Francisco Pasadas, Juan J. Palacios (2,3), Francisco G. Ruiz, Andres Godoy, Enrique G. Marin
% 1 Granada
%2 Departamento de Física de la Materia Condensada, Universidad Autónoma de Madrid, 28049 Madrid, Spain. 3 Instituto Nicolás Cabrera, Condensed Matter Physics Centre (IFIMAC), 28049 Madrid, Spain. ¶
\maketitle

\begin{abstract}
Atomic migration from metallic contacts, and subsequent filament formation, is recognised as a prevailing mechanism leading to resistive switching in memristors based on two-dimensional materials (2DMs). This study presents a detailed atomistic examination of the migration of different metal atoms across the grain boundaries (GBs) of 2DMs, employing Density Functional Theory in conjunction with Non-Equilibrium Green's Function transport simulations. Various types of metallic atoms, Au, Cu, Al, Ni, and Ag, are examined, focusing on their migration both in the out-of-plane direction through a MoS\textsubscript{2} layer and along the surface of the MoS\textsubscript{2} layer, pertinent to filament formation in vertical and lateral memristors, respectively. Different types of GBs usually present in MoS\textsubscript{2} are considered to assess their influence on the diffusion of metal atoms. The findings are compared with structures based on pristine MoS\textsubscript{2} and those with mono-sulfur vacancies, aiming to understand the key elements that affect the switching performance of memristors. Furthermore, transport simulations are carried out to evaluate the effects of GBs on both out-of-plane and in-plane electron conductance, providing valuable insights into the resistive switching ratio.
\end{abstract}

\section{Introduction}
%In the past decade \textbf{(also currently)}, there has been a significant focus on 
The current surge of data-intensive artificial intelligence (AI) applications \cite{Silver_Nature,Zhang_Drug}, %such as AlphaGo or large language models like ChatGPT. 
implementing artificial neural networks (ANNs), has an Achilles heel in the suboptimal power and latency performance of traditional von Neumann computing architectures%, as they lead to extremely high power consumption 
\cite{Milakov_Nvidia,Roy_Nature}. This so-called von Neumann bottleneck has raised enormous interest in the development of a novel neuromorphic  hardware, as a potentially transformative technology for AI \cite{Mead_VLSI,Zhu_APR}. In more detail, the crucial step for the efficient implementation of these neuromorphic systems appears to be the realization of artificial neurons and synapses capable of mimicking the essential learning rules of the biological brain, such as short- and long-term plasticity and synaptic efficacy \cite{Hebb_Wiley,Indiveri_Proced}. 

In this dawn of AI hardware, some commercial deployments, such as TrueNorth by IBM and Loihi by Intel have already demonstrated synaptic functionalities, although still utilizing traditional CMOS architectures,  \cite{Akopyan_TCAD,Davies_Micro}; while other lower readiness and more innovative proposals employ emerging structures operated using ferroelectricity or electrochemical gate coupling \cite{Fuller_IBM,Zhu_AdvMat,Jerry_IEDM}. Despite the progresses made, all these alternatives are affected by the same weakness: the use of conventional transistors, which limits their scaling capabilities. For instance, TrueNorth packs around 10\textsuperscript{6} neurons and 2.56$\times$10\textsuperscript{6} synapses \cite{Akopyan_TCAD}, which, although impressive, are far from the integration and connection capabilities of the human brain, which packs around 10\textsuperscript{11} neurons and 10\textsuperscript{15} synapses \cite{Suzana_PNAS}.
In this regard,  two-terminal memristors, with their inherent ability of event-driven state change and memory, have emerged as a highly attractive alternative for neuromorphic devices \cite{Wang_Nature,Chua_TCT}. %In addition, memristors are being explored for non-volatile memory implementation, such as resistive RAMs \cite{Wong_IEEE}.
They are composed of a layer of resistive switching material, usually based on transition metal oxides, chalcogenides, or polymers, sandwiched between two metallic electrodes. Notably, thanks to this simple architecture,  they can be manufactured in crossbar arrays, facilitating the ultra-high levels of integration required in ANNs \cite{Waser_AdvMAt,Zahoor_NRL}. %Continuing in the same line, 

Although memristor implementations have been known and developed from more than a decade ago, e.g. in non-volatile resistive RAMs \cite{Wong_IEEE}, the recent demonstration of resistive switching in 2D materials (2DMs) has put the spotlight back on them for their eventual inclusion in advanced neuromorphic devices \cite{Wang_NatElec,Hui_AdvElecMat,Huh_AdvMat}. In particular, 2DM-based memristors have, in the last few years, shown low-power and fast switching capabilities, possibility of deposition on flexible substrates, inherent ability for extreme scaling due to their atomically thin layers, and the potential to create diverse devices and functionalities due to the large library of 2DMs and their suitability for heterogeneous integration \cite{Mohit_AMI,Yang_Nanoscale,Huh_AdvMat}. Compared to amorphous oxides, the use of crystalline 2DMs offers an advantageous route to control memristive characteristics. Despite these unique features, the experimental demonstrations of 2DM-based memristors are yet to achieve their true potential, and they still exhibit a varied range of terminal characteristics. Some of the critical ones are: switching voltages from a few volts to tens of volts; operating currents from nA to mA; endurance from a few tens of cycles to a hundred of cycles; and resistance switching ratios varying from 1 to 3 orders of magnitude \cite{Hui_AdvElecMat,Huh_AdvMat}. This dissimilar performance arises because of the inherent stochastic nature of the resistance switching mechanisms and the limited understanding of how the material properties affect the resistive switching.
%Acheiveing the desirable FOMs requires that the process is understood not only qualitatively but their impact on the FOMs. 

\begin{figure*}[!t]
\centering
\includegraphics[width=0.8\linewidth]{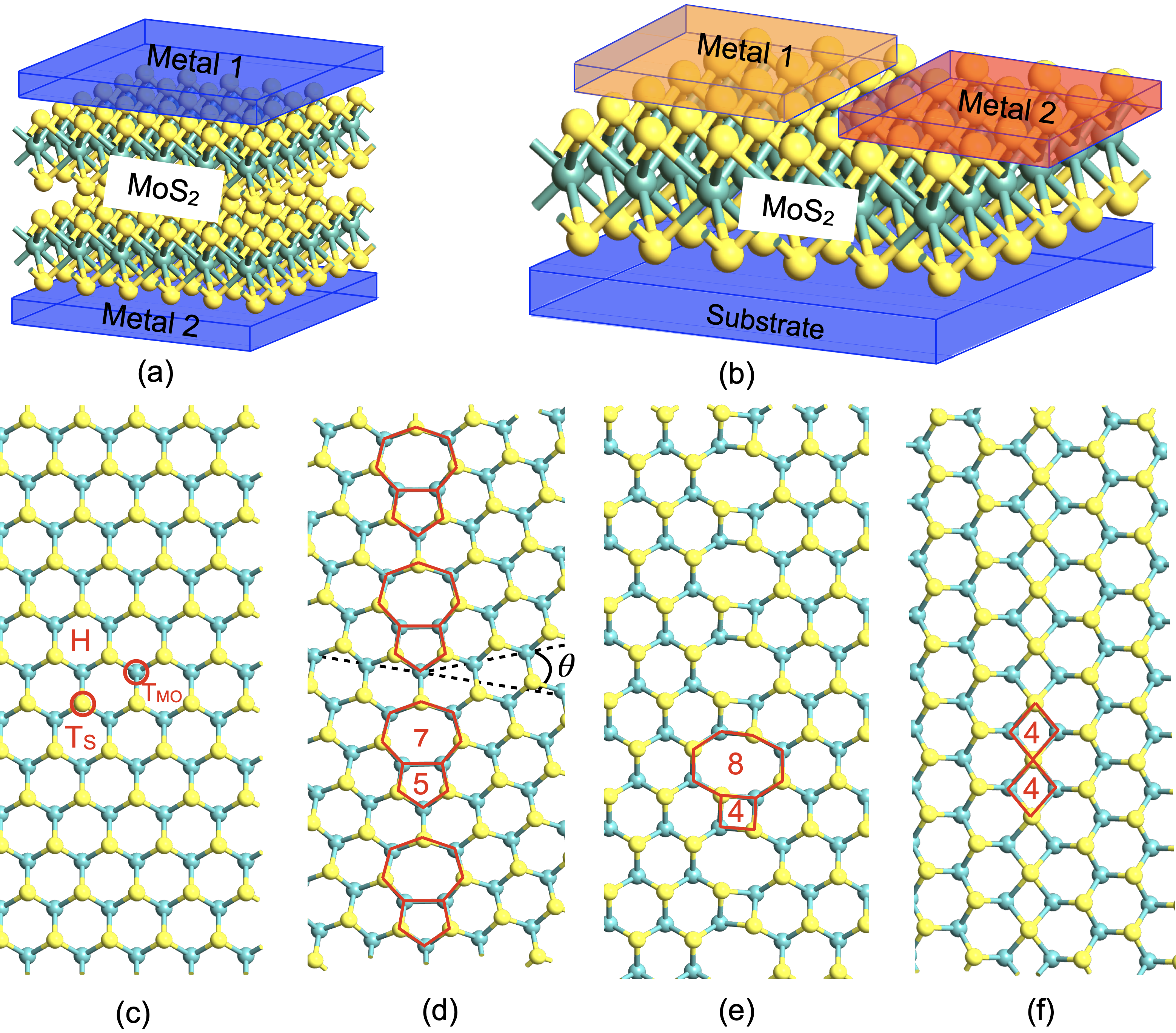}%
\label{subfig:MoS2_supercell}%
\quad \quad 
\caption{Schematic of (a) a vertical memristor and (b) a lateral memristor. Snippet of the optimized  MoS\textsubscript{2} geometries: (c) monocrystalline; (d) with a GB at a tilt angle $\theta$ (the angle of mismatch between two adjoining crystal) of 22\textdegree \ with a 5|7 dislocation core repeated continuously along a line; (e) with a zig-zag (ZZ) edge oriented GB at a $\theta$ = 60\textdegree \ resulting in a 4|8 dislocation core; and (f) with an armchair (AC) edge oriented GB at a $\theta$ = 60\textdegree \ with a 4|4 dislocation core.}
\label{fig:GB_struct}%
\end{figure*}

This divergent behavior of resistive switching (RS) in 2DMs is a consequence of their diverse structure and composition, although some common underlying phenomena can be identified. For instance, considering the simple two-terminal vertical structure depicted in Fig. \ref{fig:GB_struct}a, where a 2DM is sandwiched between a top and a bottom metal electrodes, primarily two different scenarios are encountered in the literature. If the 2DM is a monolayer, RS is attributed to the atomic point contact at the metal-2DM interface, known as the atomristor \cite{Ge_NanoLett}. On the other hand, in multilayer 2DMs, RS is mostly ascribed to one of the two main mechanisms: the formation of an interlayer bridge under an external field, as in hexagonal boron nitride (hBN) \cite{Ducry_NPJ}, or the creation of metallic filaments via atom migration from metallic contacts \cite{Xu_ACSNano,Shi_IEDM}. In the case of lateral structures, a mono or multilayer 2DM is contacted to either side by metallic electrodes, as represented in Fig. \ref{fig:GB_struct}b. In this type of devices, the RS is mainly attributed to the formation of metallic filaments through migration of metal atoms on the 2DM surface \cite{Matteo_AdvMat},\cite{Ge_AdvMat} or through the change in orientation of grain boundaries (GBs) \cite{Sangwan_Nature}. %Lateral structures based on transition metal dichalcogenides (TMDs), have shown the formation of metallic filaments as the primary mechanism for memristive switching .

The formation of metallic filaments, through the atomic jump process, is in this way pointed out as one of the prevailing mechanism for RS in 2DMs. Its dependence on various factors, including crystal structure, size, and chemical nature of the diffusing atoms, as well as whether diffusion is mediated by defects \cite{Mehrer_Book}, results in the diverse performance of experimental realizations of 2DM-based memristive devices.  %For example, in the case of more traditional RS materials such as bulk amorphous oxides, the presence of numerous interstitial sites facilitates the metal atom migration. 
Adding more complexity to the problem, the knowledge gained along the years for bulk amorphous oxides in RRAMs is not directly applicable to 2DMs: e.g. while in bulk amorphous oxides the presence of numerous interstitial sites facilitates the metal atom migration, the crystalline nature of 2DMs acts as a barrier to the movement of metallic atoms, especially in the out-of-plane direction. Additionally, the innate anisotropy of the 2D crystal makes the ion surface diffusion a distinct process to the out-of-plane one. In this context, while some studies have suggested that intrinsic monochalcogen vacancy defects in TMDs could facilitate the migration of metal atoms in the otherwise pristine structure \cite{Li_Nature,Shubhdeep_AMI}, other researchers have linked the ease of metal atom migration with the presence of dislocations and GBs in 2DMs \cite{Ge_AdvMat,Xu_ACSNano}. Recent molecular dynamics-based simulations have also demonstrated that a filament may be formed in a structure containing GBs \cite{Mitra_NPJ}. 

Although the prevailing mechanism remains unclear, the experimental evidence suggests that disorder or defects in the crystalline structure are crucial to trigger the memristive mechanism. However, in many cases, the discussion has not yet left the qualitative arena, elucidating the mechanism, but laying apart the deeper physical understanding of the process. Therefore, it becomes crucial to further analyze this issue by carefully examining the atomic nature of such defects and their relation to the migrating atoms in 2DMs. To gain insight into this transcendent issue, a detailed atomistic study can provide guiding principles in the design of optimized 2DM-based memristors, allowing for better control over the switching features and uncovering hidden phenomena that can be missed in higher abstraction-level studies. This work expands on these concepts by investigating various types of GBs commonly found in MoS\textsubscript{2} and their impact on the migration of metallic atoms through the use of Density Functional Theory (DFT) combined with Non-Equilibrium Green's Function (NEGF) transport simulations. The obtained results are compared with other structures based on pristine materials or containing a mono-sulfur vacancy. Moreover, the study also focuses on the atom migration both in the out-of-plane direction through the MoS\textsubscript{2} layer in connection to the vertical memristive structure and in the in-plane direction along the MoS\textsubscript{2} surface for the lateral memristors.

%Comapring them with other vacany defect and opristine structure. We also want to draw a connection with the FOMs and how it is related to orientation angle. To acheive this this work employed a DFT+NEGF study and NEB calculation of migration barrier 

%As compared to a bulk crystalline solids the vdW gap in the layered 2DMs provides a favourable site for the foreign adatoms. Such intercallition is already efficiently used in numerous 2DMs applicaitons. 
\section{Results and Discussion}

%Different structures simulated are shown in , which includes a pristine MoS\textsubscript{2}, and three different types of GB.

Three different types of grain boundaries (GBs)  (see Fig. \ref{fig:GB_struct}) are analyzed and compared with a pristine monocrystalline MoS\textsubscript{2} and a monocrystalline MoS\textsubscript{2} with single sulfur vacancy (V\textsubscript{S}). 
The different GBs are characterized by the value of the tilt angle ($\theta$), i.e. the relative mismatch between the orientation of the two adjoining crystal planes as represented in Fig. \ref{fig:GB_struct}d, which depicts the case $\theta$ = 22\textdegree \ with a 5|7 dislocation core repeated continuously along a line; Fig. \ref{fig:GB_struct}e, with a zig-zag (ZZ) edge oriented $\theta$ = 60\textdegree \ resulting into 4|8 dislocation core; and similarly Fig. \ref{fig:GB_struct}f, where the armchair (AC) edge oriented $\theta$ = 60\textdegree \ with 4|4 dislocation core is shown. Two of these GBs, 22\textdegree \ and 60\textdegree \ ZZ, provide a wider interstitial gap, while the 60\textdegree \ AC, results in a narrower interstitial gap compared to the pristine MoS\textsubscript{2}. The GBs are constructed using a periodic boundary condition, where two anti-parallel GBs are present in one supercell, with at least 16 \AA \ separation between them (see Supporting information Fig. S1). To ensure a fair comparison, all the supercells are constructed to be nearly the same size, resulting in a 10$\times$3 supercell of pristine MoS\textsubscript{2}. The crystals are fully relaxed with the force criteria described in the Methods Section, and the resulting structures verified through comparison with previous experimental and theoretical studies \cite{Zhou_Nanolett_exp,Zou_NanoLetters,Ping_ACR}. The diffusion barriers for ions are then evaluated using the Nudge Elastic Band (NEB) with the climbing image method \cite{Soren_JCP}, while for the calculation of the transmission coefficient in various directions the DFT + NEGF method is used, employing Landauer formalism to determine the conductance (see Methods).

\subsection{Vertical memristors}
The vertical memristor is first analyzed, where the external metallic adatom navigating in the out-of-plane direction needs to squeeze through the MoS\textsubscript{2} lattice plane. This process requires a rearrangement of the lattice atoms, raising the barrier to its diffusion. In such situations, the size of the interstitial gap provided by the lattice and the atomic radius of the adatoms become the key parameters for the atomic jump process. However, it is known that the atomic rearrangement within GBs and dislocation cores are more relaxed than in the bulk phase and therefore should provide less hindrance to the migrating species and accelerate the rate of mass transport \cite{Souza_Book,Mehrer_Book}.

\begin{figure*}[!t]
\centering
\includegraphics[width=\linewidth]{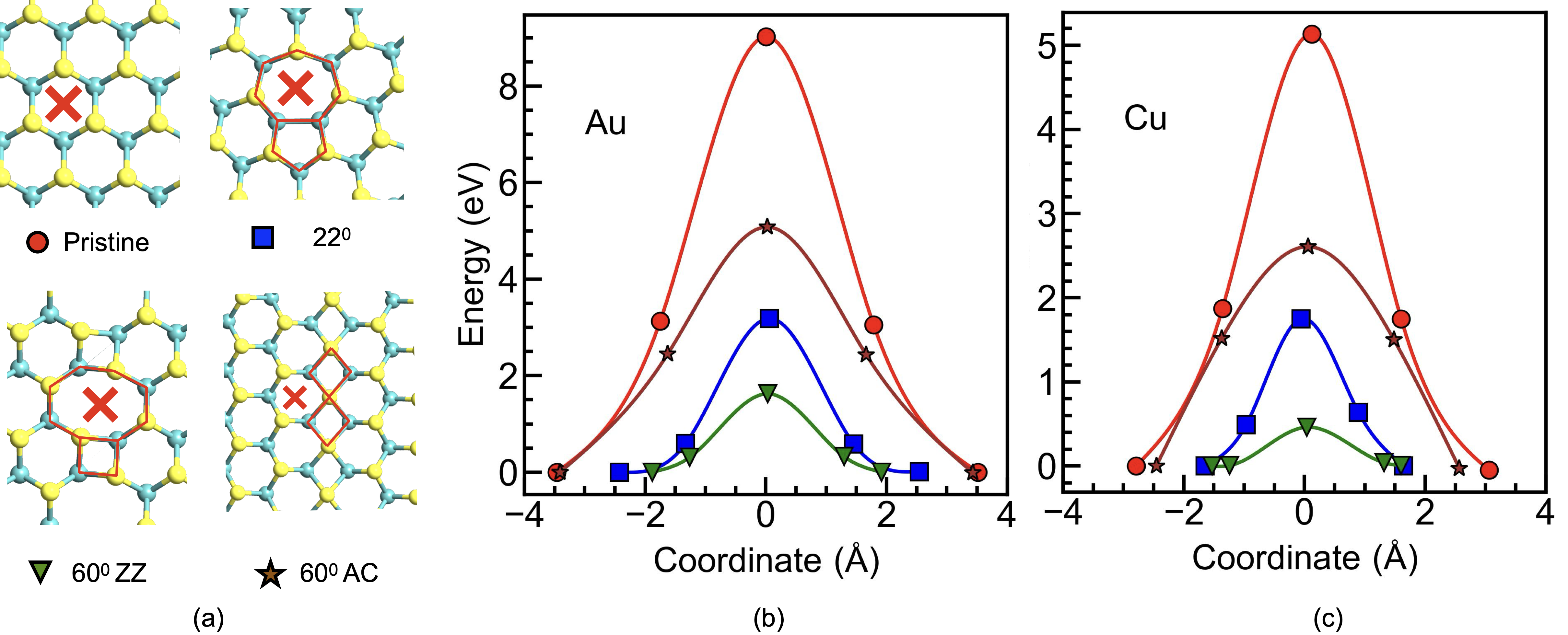}%
\label{subfig:MoS2_supercell}%
\caption{(a) Diffusion paths, (b) energies for the diffusion of Au, and (c) Cu in the out-of-plane direction calculated for a pristine MoS\textsubscript{2} and MoS\textsubscript{2} with $\theta = $ 22 \textdegree, 60 \textdegree \ ZZ and 60 \textdegree \ AC grain boundaries.}
\label{fig:NEB_all_OOP}%
\end{figure*}

To calculate the activation energy required for the metal adatom to diffuse perpendicular to the MoS\textsubscript{2} layer, the adatom is considered at the octahedral (H) site, shown by the cross mark in Fig. \ref{fig:NEB_all_OOP}a for pristine MoS\textsubscript{2}, and also for those sites with the highest interstitial gap in the GBs (see Fig. \ref{fig:NEB_all_OOP}a). The calculations are carried out first for Gold (Au) and Copper (Cu), as they respectively represent a noble and an active metal widely used as contact electrode in memristors. The NEB calculation requires to know the initial and final position of the diffusing atom, and therefore, both locations are found by fully relaxing the structure with the metal atom on top of the MoS\textsubscript{2} monolayer (initial) and the metal atom at the bottom of the MoS\textsubscript{2} layer (final). Figures \ref{fig:NEB_all_OOP}b-c show the energy variation along the minimum energy path (MEP) for both metal adatoms and the different GB configurations. Note that here the diffusion barrier energies are plotted against the distance traveled by the adatom in the out-of-plane direction, where the coordinate 0 is a reference position corresponding to the plane of the Mo atoms in the S-Mo-S atomic sandwich of MoS\textsubscript{2}, signifying the location of the mono-layer. The first observation that must be made is the different initial and final position of the Au and Cu adatoms with respect to the MoS\textsubscript{2} monolayer, which can be noted as d\textsubscript{Au-Mo} and d\textsubscript{Cu-Mo}, respectively. The d\textsubscript{Au-Mo} (d\textsubscript{Cu-Mo}) is 3.46 (3.41) \AA \ for the case of pristine MoS\textsubscript{2}, which reduces to 2.43 (1.64) \AA \ and 1.88 (1.53) \AA \ in the case of the 22\textdegree \ and 60\textdegree \ ZZ GBs, respectively. This decrease in the Au-Mo gap due to the presence of the GB agrees well with other reported values \cite{Gabriele_ANM} and has an important influence on the contact resistance by reducing the tunneling barrier width . Regarding the diffusion barrier, our calculations reveal that Au (Fig. \ref{fig:NEB_all_OOP}b) needs to surmount an extremely high barrier of 9.02 eV for diffusing through the pristine MoS\textsubscript{2} without any defects (red circles). However, compared to the pristine case, both the 22\textdegree \ and 60\textdegree \ ZZ GBs (blue squares and green triangles) reduce the diffusion barrier considerably, resulting in a value of 3.19 eV and 1.62 eV, respectively. Similarly, as shown in Fig. \ref{fig:NEB_all_OOP}c, pristine MoS\textsubscript{2} provides the highest barrier for Cu, followed by the MoS\textsubscript{2} with 22\textdegree \ and 60\textdegree \ ZZ GBs. It is also evident that the higher interstitial gap provided by the 8-fold ring in the 4|8 dislocation appearing in the 60\textdegree \ ZZ GB (see Fig. \ref{fig:NEB_all_OOP}a) provides the lowest hindrance to the movement of external metal atoms. The calculated diffusion barriers are listed in Table \ref{tbl:Barrier_OOP}. The case of the MoS\textsubscript{2} with 60\textdegree \ AC GBs, however, presents both for Au and Cu a rather different behavior. Using the same force criteria, the NEB calculations was unable to find any converged MEP for external metal atom movement through the 4|4 dislocation, suggesting a higher difficulty for the atom to move through the smaller interstitial gap. Instead, the barrier through the adjacent hexagonal ring (shown by the cross mark in Fig. \ref{fig:NEB_all_OOP}a) is computed and plotted in Fig. \ref{fig:NEB_all_OOP}b-c (brown stars), and it is found that the barrier is reduced compared to the pristine case for both Au and Cu, although still with higher diffusion barriers than the ZZ configurations. Although this hexagonal ring is similar to the one of the pristine situation, the proximity of the GB results in stretched bonds, thereby increasing the interstitial gap. 

Finally, there is the MoS\textsubscript{2} with a mono-sulfur vacancy (V\textsubscript{S}), which presents a substitutional site for the metal atom, as also reported in \cite{Ge_AdvMat}. The structure with V\textsubscript{S} was created by removing a single sulfur atom (T\textsubscript{S} site) from the supercell of the pristine MoS\textsubscript{2} shown in Fig. \ref{fig:GB_struct}c. In this scenario, no converged MEP for the out-of-plane diffusion could be found, suggesting that the metal atoms remained trapped in the V\textsubscript{S}. This result obtained here also explain the absence of filament formation in the presence of sulfur vacancies in the molecular dynamics study carried out in \cite{Mitra_NPJ}.
\begin{table}[!t]
\small
  \caption{\ Diffusion barrier energies for different external metallic adatoms diffusing out-of-plane through a single crystalline MoS\textsubscript{2} monolayer, MoS\textsubscript{2} with mono-sulfur vacancy (V\textsubscript{S}) and GBs with a tilt angle of 22\textdegree, 60 \textdegree \ ZZ, 60\textdegree \ AC, and 13.16 \textdegree. Aluminum atom prefers to get absorbed in the middle of the 8-fold ring for 60 \textdegree \ (ZZ) GB, presenting a local minimum there. $\dag$ barrier is calculated at the adjacent site of the 4|4 dislocation ring.}
  \label{tbl:Barrier_OOP}
  \begin{tabular*}{0.48\textwidth}{@{\extracolsep{\fill}}cccccl}
    \hline
    \multirow{2}{*}{Type} & \multicolumn{5}{c}{Barrier (eV)} \\
                      & Au   & Cu   & Al   & Ni   & Ag   \\ \hline
pristine              & 9.02 & 5.13 &     4.32 &    4.04  &   8.96   \\ \\
V\textsubscript{S}                    & -  & -  &   -   &   -   &   -   \\ \\
22\textdegree                   & 3.19 & 1.75 & 0.81 & 1.51 & 3.72 \\ \\
60\textdegree \;ZZ                 & 1.62 & 0.46 &   -   &   0.35   & 1.59     \\ \\
60\textdegree \; AC$^\dag$ & 5.08 & 2.6 & 1.7 &2.87 & 5.18 \\ \\
13.16\textdegree    & 2.73  & 1.59& 0.55 & 1.27 & 3.41 \\
    \hline
  \end{tabular*}
\end{table}

The calculations reveal that, in all scenarios, the diffusion barrier for copper (Cu) is lower than that for gold (Au), a result that aligns with expectations given Cu smaller atomic radius. To further investigate, the Minimum Energy Path (MEP) for additional metals commonly used as electrodes in memristors, which vary in atomic radius and chemical reactivity, was computed (see Table \ref{tbl:Barrier_OOP}). Notably, all investigated metals follow the same dependence with the GB configuration facing the highest barrier for the pristine material followed by the 60 \textdegree \ AC GB; and with decreasing barriers for 22 \textdegree, 13.16 \textdegree \ GBs, and the smallest diffusion barrier observed for the 60 \textdegree ZZ GB. A comprehensive description of the energy barrier along the MEP for different GB structures is provided in the Supporting information Fig. S2 for Al, Ni, and Ag.

\begin{figure}[!t]
\centering
\includegraphics[width=\columnwidth]{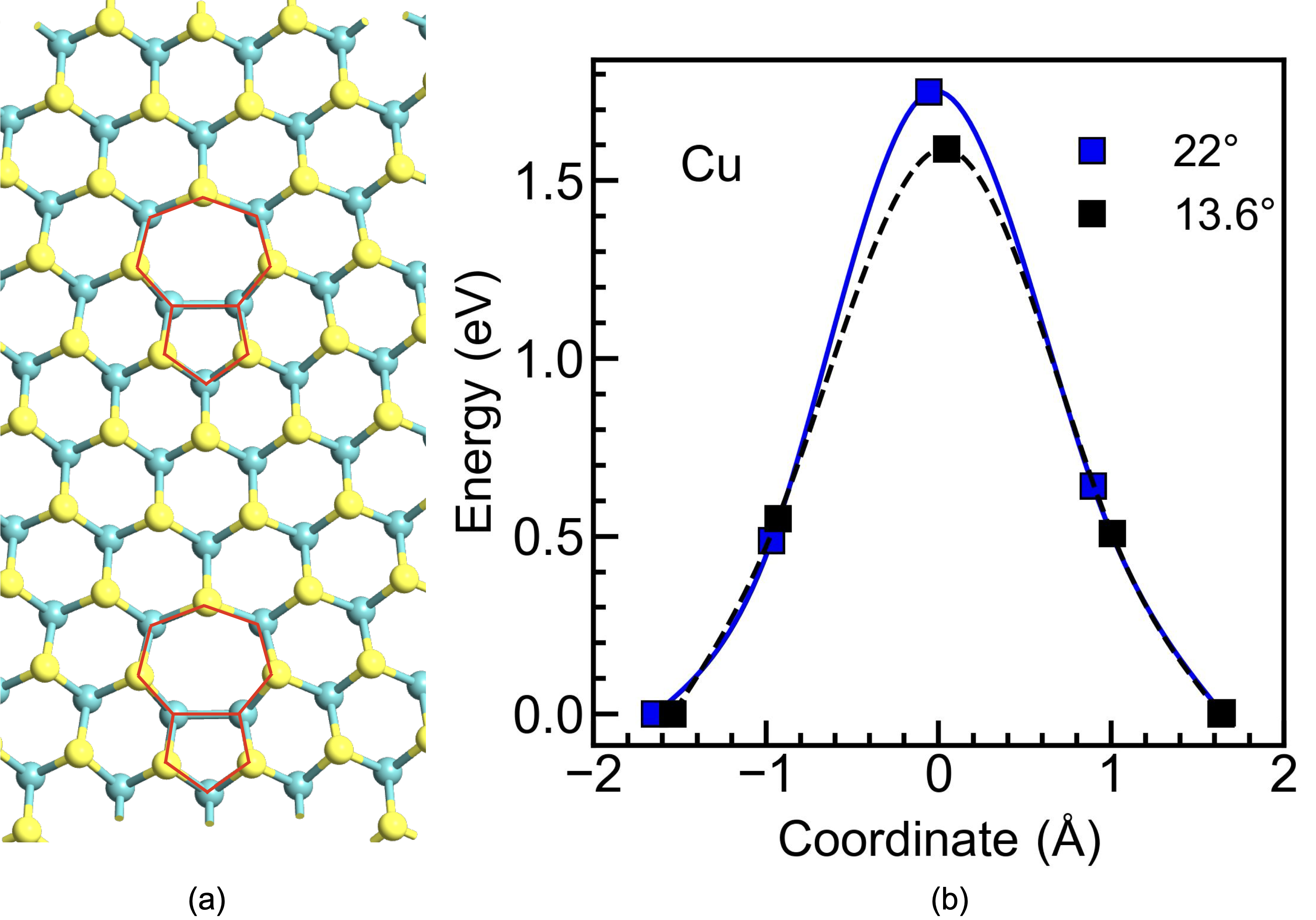}%
\label{subfig:MoS2_supercell}%
\quad \quad 
\caption{(a) Snippet of the supercell of monolayer MoS\textsubscript{2} with a $\theta =$ 13.16\textdegree \ GB where the 5|7 dislocation cores (highlighted with red line) are localized and separated from each other and (b) Calculated energies for the diffusion of Cu in the out-of-plane direction for MoS\textsubscript{2} with $\theta = $ 22\textdegree and 13.6\textdegree grain boundaries. }
\label{fig:NEB_1e}%
\end{figure}

To further examine the influence of the proximity of the dislocation cores on the observed diffusion barrier, a low-angle GB of $\theta$=13.16\textdegree \ was constructed, which also results in 5|7 dislocation cores. Nevertheless, these are localized and separated from each other (see Fig. \ref{fig:NEB_1e}a) as opposed to the $\theta$=22\textdegree \ case, in which the 5|7 cores are continuously repeated (see Fig. \ref{fig:GB_struct}d). The diffusion barriers calculated in both scenarios are quite similar, as indicated in Fig. \ref{fig:NEB_1e}b, suggesting that their value is governed by the localized interstitial gap, and not by the overall GB structure. As the barrier of the nearby pristine H-site will be much higher than the one in the 7-fold ring, this will result in localized sites with preferential filament formation. This could be further pursued as a way to preserve the crystalline nature of the pristine MoS\textsubscript{2} with a few localized sites for filament formation, rather than a continuous GB. 

In order to relate the microscopic diffusion barrier to the macroscopic performance, it could be thought that higher external forces, such as an increased electric field or temperature, would be required to overcome a higher barrier, thus influencing the measured switching voltage ($V$\textsubscript{switch}) of the memristors. Therefore, our previous calculations suggest that the presence of grain boundaries becomes crucial to lower the switching voltage of the memristors, as desired for low-power operation. At the same time, the continuous repeating dislocation core occurring in the GBs will provide multiple sites with similar energy barrier to diffuse and, therefore, similar probability of filament formation. This feature would result in the potential formation of multiple filaments, increasing the variability during continuous switching events. On the contrary, the presence of a lower angle GB can provide a compromise, where the number of sites for filament formation will be reduced and consequently the device variability. Furthermore, from the above results, the choice of the metal electrode plays a critical role, not only in reducing the tunneling barrier between the contact and the 2D material, but also in minimizing $V$\textsubscript{switch} for those adatoms showing lower diffusion barriers for the same GB. \\

\subsection{Lateral memristors}
\begin{figure*}[!t]
\centering
\includegraphics[width=\linewidth]{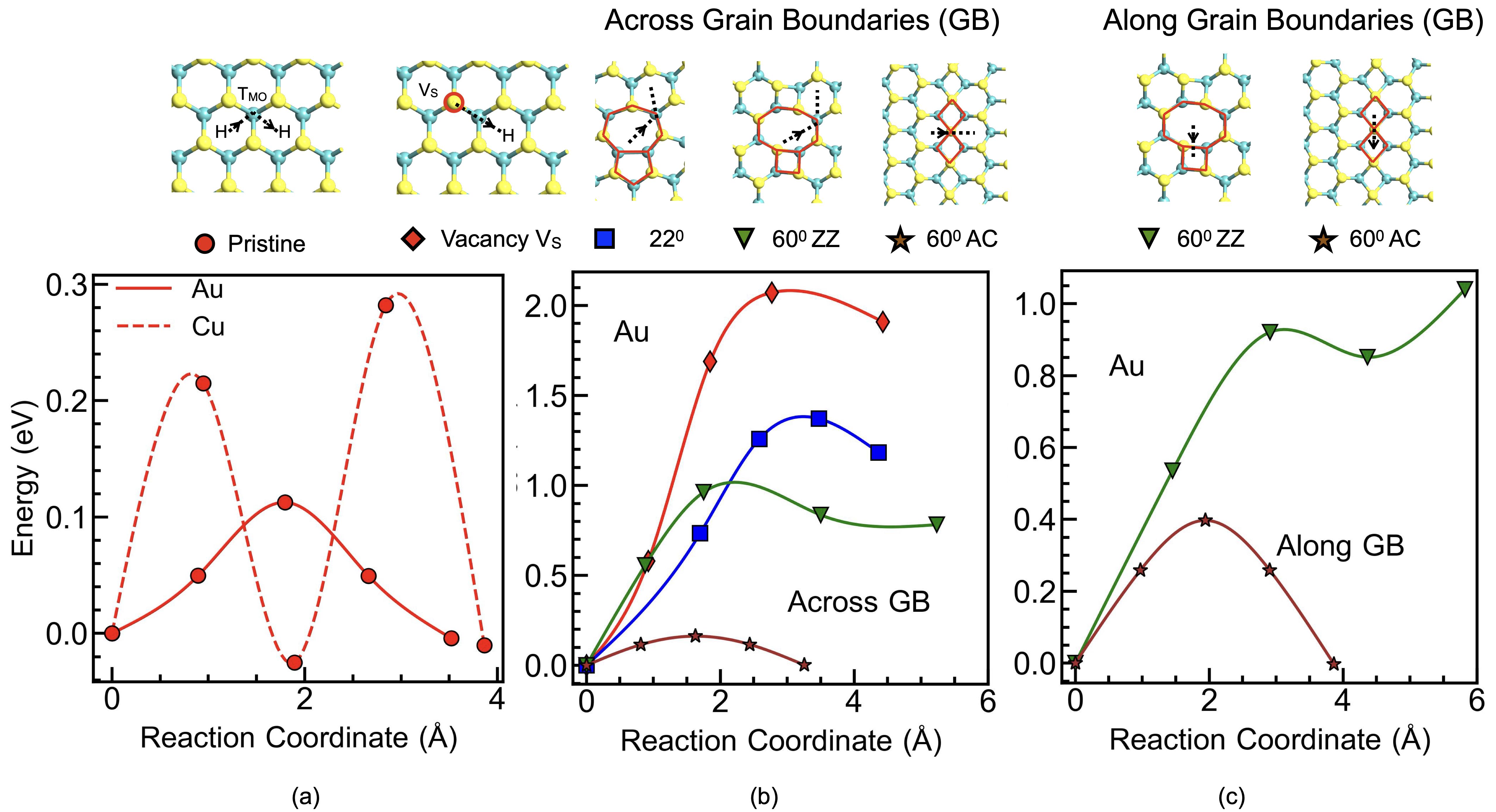}%
\label{subfig:MoS2_supercell}%
\caption{Calculated energies for the diffusion of (a) Au and Cu on the surface of a single crystalline MoS\textsubscript{2}, (b) Au on the surface of MoS\textsubscript{2} with V\textsubscript{S}, and across the grain boundaries with $\theta = $ 22\textdegree, 60\textdegree \ ZZ and 60\textdegree \ AC and (c) Au along the grain boundaries with $\theta = $ 60\textdegree \ ZZ and 60\textdegree \ AC. The diffusion paths for each case are indicated on the top panel.}
\label{fig:NEB_all_IP}%
\end{figure*}

Next, the diffusion of the metallic atoms on the surface of the MoS\textsubscript{2} layer is analyzed, a process needed to form a conductive path in the lateral memristors. Figure \ref{fig:NEB_all_IP}a illustrates the diffusing path on the surface of pristine MoS\textsubscript{2} (top panel) along with the activation energy required for both Au and Cu, indicating the movement from one H-site to the adjacent H-site via a Mo-tetrahedral site (T\textsubscript{Mo}). It can be seen that Au (solid lines) experiences a barrier of only 0.11 eV, whereas Cu (dashed lines) needs to overcome a barrier of around 0.307 eV. The surface diffusion barriers calculated here are in good agreement with previous theoretical calculations \cite{Guzman_JAP,Ping_PCCP}. It is important to note that the minimum energy site for the adsorption of the metallic adatoms on the MoS\textsubscript{2} surface could vary among different metals. For instance, from Fig. \ref{fig:NEB_all_IP}a, one can infer that Au prefers the H-sites whereas the Cu has lower energy at the T\textsubscript{Mo} ones. Indeed, different diffusion paths could be considered for each atom \cite{Guzman_JAP,Ping_PCCP, Sun_SR,Shi_MCP}; %However, in order to be consistent 
here, the diffusion path has been taken to be from one H-site to the nearest H-site. Nevertheless, since just the maximum and the minimum energies along the diffusion path are meaningful for the definition of the barrier, the same value  would result in the case of defining the diffusion path from one T\textsubscript{Mo}-site to the nearest T\textsubscript{Mo}-site. From the results, it can be concluded that the barrier for the surface displacement (i.e. along the MoS\textsubscript{2} plane) is significantly lower compared to the out-of-plane one (i.e. through the MoS\textsubscript{2} layer). This is consistent with the fact that the pristine 2D surface is free from dangling bonds, facilitating the diffusion of the adatoms. However, an interesting observation is made when the diffusing atom encounters a V\textsubscript{S} or a GB. As aforementioned, the V\textsubscript{S} behaves as a trapping site for diffusing adatoms, and this behavior is in-fact responsible for the formation of an Au-point contact in the case of the atomristor \cite{Ge_AdvMat}. However, as lateral memristors are based on the formation of a filament connecting two lateral contacts, the V\textsubscript{S} acts as a roadblock for the diffusion of metallic atoms. As depicted by red diamonds in Fig. \ref{fig:NEB_all_IP}b, an extra energy of around 2.07 eV is required to pull the Au-atom out from the V\textsubscript{S} to the adjacent pristine H-site. 

Similarly, two scenarios were tested for the GBs. First, the Au atom diffusing across the GBs, i.e. from the top of the dislocation core (either 22\textdegree \ --blue square-- or 60\textdegree \ ZZ --green triangles--) to the nearest pristine H-site, see top panel of Fig. \ref{fig:NEB_all_IP}b. In this scenario, it is found that the higher ring in the dislocation, which provided the lowest barrier for out-of-plane diffusion, acts as trapping site for surface diffusion. The calculation reveals that Au experiences a barrier of around 1.37 eV and 0.96 eV for diffusion across the 22\textdegree \ and 60\textdegree \ ZZ  tilted GB, respectively. Similar results are found for Cu (see Table \ref{tbl:Barrier_IP} and the Supporting information Fig. S3). In other words, the 7-fold and 8-fold rings occurring in the GBs with $\theta = $ 22\textdegree \ and 60\textdegree \ ZZ are energetically more favorable for the Au atom than the adjacent pristine H-site, suggesting that (similar to the V\textsubscript{S}) the GBs also act as trapping sites for adatoms. This fact is aligned with the previously obtained lower values of d\textsubscript{Au-Mo} at the GB compared to the pristine H-site, proving their stronger interaction. Such a preferential location for the adatom adsorption provided by the GBs can be useful for valuable applications such as catalysis or for reducing the contact resistances, but it acts as a roadblock to the lateral filament formation required for the memristive operation. However, the 60\textdegree \ AC tilted GB (brown stars in Fig. \ref{fig:NEB_all_IP}b top panel) has two symmetric locations across the GB, corresponding to the H-sites of the two stretched neighbouring hexagon-rings, providing an energetically favourable position for the Au atom and showing a barrier of just 0.16 eV to cross the GB, much lower than the other two GBs and closer to the pristine MoS\textsubscript{2}. All the calculated barriers are listed in Table \ref{tbl:Barrier_IP}.

\begin{table}[!t]
\small
  \caption{\ Diffusion barrier energies (in eV) for different external metallic adatoms diffusing on the surface of a single crystalline MoS\textsubscript{2} monolayer, MoS\textsubscript{2} with mono-sulfur vacancy (V\textsubscript{S}) and grain boundaries with tilt angle of 22\textdegree, 60\textdegree \ AC and 60\textdegree \ ZZ.}
  \label{tbl:Barrier_IP}
  \begin{tabular*}{0.48\textwidth}{@{\extracolsep{\fill}}ccccl}
    \hline
    \multirow{2}{*}{Type} & \multicolumn{4}{c}{Barrier (Across/Along) (eV)} \\
                      & Au   & Cu   & Al   & Ag   \\ \hline
Pristine              & 0.11 & 0.307 & 0.2374     &  0.064    \\ \\
V\textsubscript{S}                    & 2.07  &  1.2   &   0.76   &  1.52    \\ \\
22\textdegree                   & 1.37 & 1.01 & 1 & 1.10\\ \\60\textdegree \ ZZ                 & 0.96/1 & 0.92/1.26&   1.4/1.5   &  0.28/0.43    \\ \\
\multicolumn{1}{l}{60\textdegree \ AC} & \multicolumn{1}{l}{0.16/0.4} & \multicolumn{1}{l}{1.15/0.68}  & \multicolumn{1}{l}{0.23/0.18} & 0.54/0.36 \\ \\
    \hline
  \end{tabular*}
\end{table}

Next, in the scenario where the movement is along the direction of the GBs (top panel of Fig. \ref{fig:NEB_all_IP}c), the diffusion barrier shows similar trends: the 60\textdegree \ ZZ  GB exhibits the same trapping behavior, where the 8-fold ring in the 4|8 dislocation core (green triangle) provides the preferred location for adsorption of adatoms, and the 60\textdegree \ AC  GB (brown stars) presents a lower barrier.  The scenario for the  diffusion of Au along the GB with $\theta = $ 22\textdegree \ is more involved, as no local minima exist at the site on  top of the 5-fold ring. At this point, the Au atom is pushed to the adjacent site on top of the 7-fold ring during structural relaxation of the initial position, suggesting even stronger trapping behavior for the atom to move along the $\theta = $ 22\textdegree \ GB. In almost all cases, a common observation is that mono-crystalline MoS\textsubscript{2} with pristine surface offers the lowest barrier to surface diffusion. Looking further for the pristine surface, it is observed that Cu presents a higher barrier than Au, an opposite trend to the situation in the out-of-plane diffusion (see Table \ref{tbl:Barrier_IP}). Further testing of Al and Ag surface diffusion on pristine MoS\textsubscript{2} indicates that Cu needs the highest activation energy, followed by Al and Au (see Table 2), whereas Ag requires the lowest activation energy, with a value of just 0.064 eV. The trend however, varies depending on the species of the metal atom for V\textsubscript{S} and GB (all the MEP for surface diffusion of the aforementioned atoms are provided in Supporting information  Fig. S3). However, unlike out-of-plane diffusion in MoS\textsubscript{2}, where the GBs can reduce $V$\textsubscript{switch} of vertical memristors, here they have the opposite effect and can even increase the V\textsubscript{switch} value for lateral diffusion. Therefore, the mono-crystalline pristine structure seems to be the preferred choice for achieving low-voltage switching in lateral memristors. Additionally, the optimal choice of metals for the electrodes in lateral memristors turns out to be different from that for vertical ones. In the case of vertical memristors, Al or Cu show the lowest energy barrier for diffusion, whereas Au or Ag appear to be appropriate electrode materials for achieving low-voltage switching in lateral memristors.
%Therefore, surface diffusion is opposite to out-of-plane diffusion.

%Linking the diffusion barrier to the macroscopic quantity, a higher barrier will likely require a higher external field to overcome it, affecting the switching voltage of the memristors. The choice of the metal electrode as well as the structure of the GB will play a crucial role, specially in vertical memristor the GB can help to reduce the switching voltage substantially. 

\begin{figure*}[!t]
\centering
\includegraphics[width=1\linewidth]{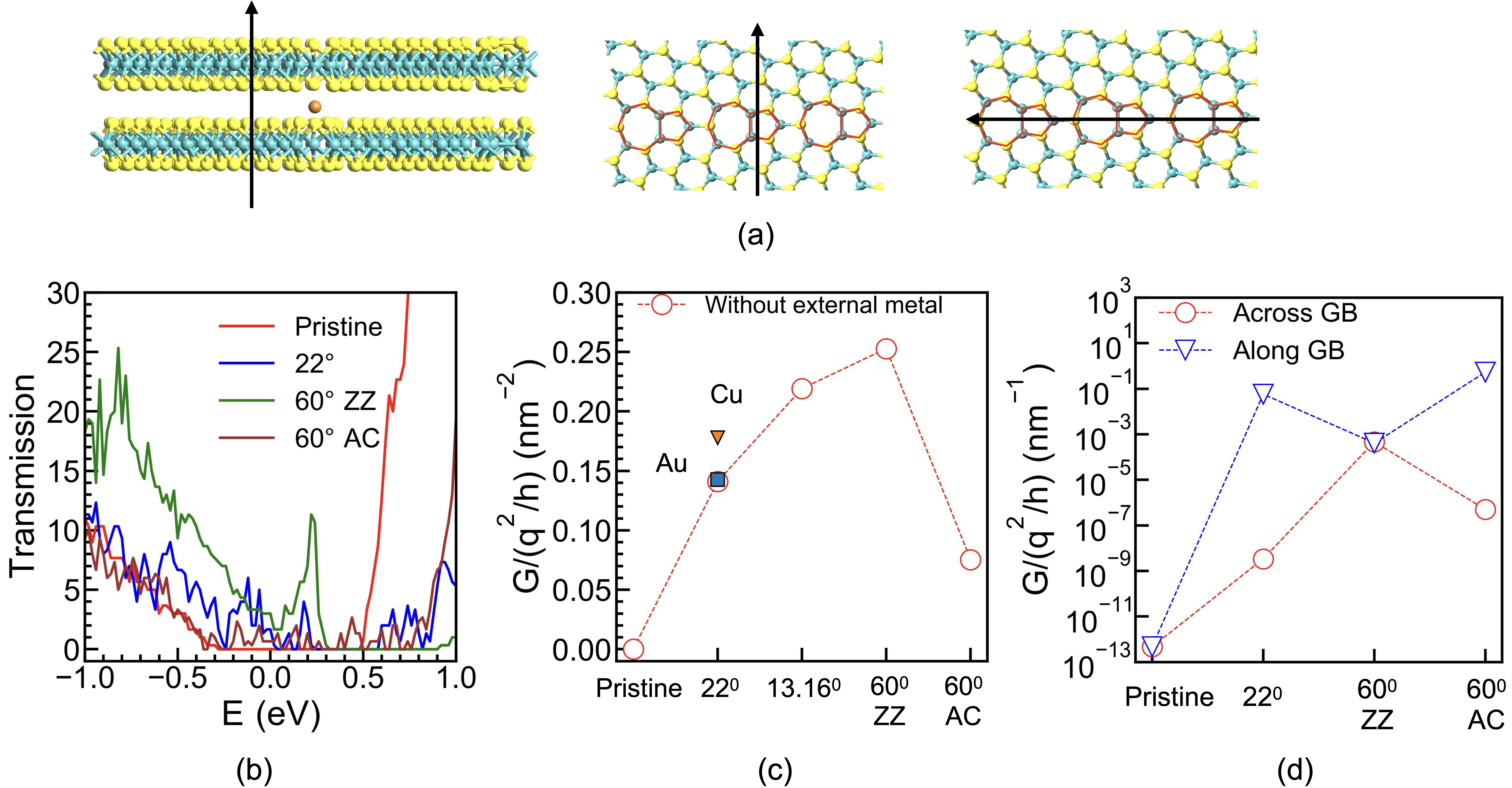}%
\quad \quad 
\caption{(a) Schematic depicting of the direction of electron transport, (b) Transmission spectrum for the out-of-plane transport through a bulk pristine MoS\textsubscript{2}, and MoS\textsubscript{2} with different grain boundaries, (c) Calculated normalized conductance for bulk pristine MoS\textsubscript{2} and with various grain boundaries, without the presence of external metal atoms (open symbols) and with
Au and Cu intercalated in the vdW gap (closed symbols) and
(d) Calculated normalized conductance for the in-plane transport in a mono-layer pristine MoS\textsubscript{2} and for MoS\textsubscript{2} with GBs, where the transport along the GBs are shown with open circles and across the GB with open triangles.}
\label{fig:Tx}%
\end{figure*}

\subsection{Electron transport}
Finally, the impact of GBs on electron transport is analyzed for both out-of-plane and in-plane scenarios. Fig. \ref{fig:Tx}a sketches both transport directions, together with the crystal arrangements. For out-of-plane, the transport through the GB is considered both in absence and presence of a metal adatom within the van der Waal gap and compared with the pristine case. The atom is placed in the most energetically favorable position according to the geometry optimization (see Methods). 
Figure \ref{fig:Tx}c shows the corresponding transmission spectrum for the out-of-plane transport with the pristine MoS\textsubscript{2} and different types of GBs, without the presence of any external metal atoms. Note that a bulk supercell is considered for calculating the out-of-plane transport. The conductance (normalized with the cross-sectional area of the supercell), extracted using Landauer formalism (see Methods), is presented in Fig. \ref{fig:Tx}d again for the pristine MoS\textsubscript{2} and with different types of GBs. Closed symbols report on the scenario for MoS\textsubscript{2} with metal atoms (squares for Au, triangles for Cu) intercalated in the vdW gap while empty symbols correspond to the case in absence of any external metal atom. The extracted conductance shows that the pristine monocrystalline MoS\textsubscript{2} has negligible out-of-plane conductance, with a normalized value in the order of $\sim$ 10\textsuperscript{-7} (nm\textsuperscript{-2}). However, the presence of GBs introduces a higher transmission, suggesting an increase in the conductivity of the multilayer MoS\textsubscript{2} as high as 5 orders of magnitude even without any external metal atom present in the crystal. Although the conductance values are quite similar for different GBs, the 60\textdegree \ AC reveals the lowest conductivity among the GBs, with increasing values by the 22\textdegree, 13.16\textdegree, and 60\textdegree \ ZZ, respectively. Interestingly, the insertion of a metal atom in the vdW gap of the structure containing GBs does not produce a noticeable effect on the conductivity. This conclusion was tested for the 22\textdegree \ tilted GB, where the initial normalized conductivity value of 0.14 nm\textsuperscript{-2} changes to 0.143 nm\textsuperscript{-2} and 0.18 nm\textsuperscript{-2}, respectively, in case a single Au or Cu atom is intercalated in the vdW gap of the supercell.
 
 In the context of a memristor exploiting the GB structure, it would be advisable that the switching  layer showed a lower conductance when no metallic filament is formed, that is so to establish the high resistance state (HRS) of the device. This layer should then be switched, by the insertion of external metal atoms through the GB, to a low resistance state (LRS); nevertheless, the observations evidence that the increase in the conductance in this latter case is not substantial.
 Thus, while GBs appear to be crucial for reducing $V$\textsubscript{switch} as previously mentioned, they also result in an unsuitable high conductance even in absence of external metal atoms (i.e. in the  HRS), thereby shrinking the resistance switching ratio between the HRS and LRS, an important figure of merit for memristor operation.
However, a caveat should be mentioned, as these simulations reflect a reduced distance between the GBs due to the limited size of the supercell. The grain sizes in the experiments may be larger. Therefore, while the alteration of the conductance originated by GBs will indeed influence the HRS, the exact quantitative change may differ from the findings reported here. Furthermore, the calculation of the switching ratio is based on the intercalation of a single metallic atom in the vdW gap of a supercell, whereas actual memristors may have multiple atoms forming a filament; nevertheless, incorporating more atoms in the vdW gap or forming a filament does not substantially change the conductance, demonstrating a saturating behavior as observed in the molecular dynamic simulations \cite{Mitra_NPJ}. Therefore, while GBs may reduce the switching voltage, they could originate a trade-off in the maximum value of the HRS and, therefore, on the switching ratio.

As for the in plane transport, as sketched in Fig \ref{fig:Tx}a, two scenarios have been considered: transport along and across the GB. The presence of GBs has an even stronger impact in this case, increasing the in-plane conductance by many orders of magnitude (with respect to the pristine case), as shown in Fig. \ref{fig:Tx}e for both the conductance along the GB (open triangles) and across the GB (open circles) -- see Fig. \ref{fig:Tx}a for the direction of transport. Note, that here the conductance is normalized with the cross-sectional length perpendicular to the direction of transport. The corresponding transmission spectrum is collected in the Supporting information Fig. S4. Although the overall conductance with GBs is higher in both directions than in the single-crystalline case, it is evident from the calculations that altering the orientation of the GBs with respect to the carrier flow can lead to a significant modulation of the conductance. For example, for $\theta = $ 22\textdegree \, the ratio between transport across and along the GB is approximately $\approx$ 10\textsuperscript{8}. Moreover, contrary to the out-of-plane situation, the change in conductance with the tilt angle is also substantial (see the 22\textdegree \ and 60\textdegree \ ZZ GBs). This orientation-dependent conductance change has been experimentally demonstrated as one of the possible switching mechanisms for the realization of a lateral memristor \cite{Sangwan_Nature}. Nevertheless, note that, in a real poly-crystal, carriers may encounter randomly oriented GBs and thereby experience both scenarios simultaneously. Besides utilizing GBs themselves as a mean for resistance tunning, the presence of GBs in a lateral memristor does not seem to favor either the formation of a filament as previously discussed or to achieve a lower current in the HRS with respect to the pristine case.\\

%The detailed features of the atomic jump process depend on various factors such as crystal structure, size and chemical nature of the diffusing atom, and whether diffusion is mediated by defects or not. 
%The crystal lattice restricts the positions and the migration paths of atoms and allows a simple description of each specific atom displacements. This contrasts with a gas, where random distribution and displacements of atoms are assumed, and with liquids and amorphous solids, which are neither really random nor really ordered. (diffusion in solids)

%The main text of the article\cite{Mena2000} should appear here.

\section*{Conclusions}
An atomistic study was conducted on MoS\textsubscript{2} to investigate the impact of structural defects, specifically different types of grain boundaries, on the main features of memristive devices. The study examined the migration of various external metal atoms both in the out-of-plane direction through the MoS\textsubscript{2} layer, which is related to the filament formation in vertical memristive structures, and on the surface of MoS\textsubscript{2}, related to the filament formation in lateral memristors. It was found that in the first scenario, the pristine structure presents an extremely high barrier for diffusion, while the presence of grain boundaries can substantially reduce this barrier, giving rise to a lower switching voltage. It was also observed that the localized dislocation is the governing factor for the diffusion barrier. Therefore, the low angle grain boundaries that generate more infrequent dislocations can possibly reduce the device variability by providing fewer localized sites for filament formation while preserving the pristine structure for most of the material. However, GBs also lead to increased transport in the out-of-plane direction, resulting in a trade-off between reducing the switching voltage and achieving an acceptable switching ratio. Regarding lateral memristors, the presence of GBs in the structure does not favor either the formation of a filament or the reduction of the current in the high resistance state. Finally, for both scemarios, the external metal atom was also found as a crucial factor in determining the memristive characteristics; in particular, Al or Cu displayed the lowest energy barrier for diffusion in vertical memristors, while Au or Ag appears
to be appropriate electrode materials for achieving low-voltage switching in lateral memristors.
%({\color{red} Therfore...}

\section{Methods}

\subsection*{Density Functional Theory}
DFT calculations are carried out using the Generalized Gradient Approximation (GGA), as implemented in QuantumATK, with a Linear Combination of Atomic Orbitals (LCAO) basis using the Perdew-Berke-Erzenhof exchange-correlation functional \cite{QATK}. SG15 pseudo-potential with medium size basis set is considered for the geometry optimization. Brillouin-zone integration was performed over a grid of k-points with a density (\AA) of 4$\times$4$\times$1 in the Monkhorst-Pack scheme. An energy cut-off of 150 Rydberg was taken. To avoid any interaction between the periodic images of the neighboring slabs in monolayer structures, a vacuum region of at least 20 \AA \ is included in the out-of-plane direction. The van der Waals (vdW) interactions are taken into account in the calculations through Grimme's DFT-D2 dispersion corrections. The grain boundary structures were generated by rotating the adjoining crystal with appropriate angles using Atomsk tool \cite{Atomsk}. The resultant geometries and lattice parameters were fully relaxed using Quantum ATK until the forces acting on each atom were $<$ 0.01 eV/\AA. 
%\textbf{Introduce the structures to be simulated more softly. Include more explanations}

%We have investigated three different types of grain boundaries  (GB) structures (see Fig. \ref{fig:GB_struct}) comparing them with a pristine mono-crystalline MoS\textsubscript{2}, a mono-crystalline MoS\textsubscript{2} with single sulfur vacancy (V\textsubscript{S}). 

\subsection*{Nudge Elastic Band}
The diffusion barrier is obtained using the Nudge Elastic Band (NEB) with the climbing image method \cite{Soren_JCP}. For NEB image optimization, the FHI pseudopotential with a Double Zeta Polarized (DZP) basis set and force criteria of less than 0.05 eV/Å is considered to optimize computational time. The accuracy of the DZP pseudopotential is verified by comparing the electronic band-structure of GB with that obtained from SG15.

\subsection*{Transport Calculations}
To calculate the transmission coefficient in various directions, we have made use of the DFT+NEGF method. A bulk structure, periodic in the $x$, $y$, and $z$ directions, is used for out-of-plane transport, while a monolayer periodic in the $x$ and $y$ directions is used for in-plane transport analysis. For the DFT calculation a DZP pseudopotential with PBE exchange-correlation function, an energy cutt-off of 150 Rydberg and $k$-point density (\AA) of 4$\times$4$\times$4 in the Monkhorst-Pack scheme is used. As for the NEGF transport, the $k$-point density (\AA) set in the direction normal to the transport is 7$\times$7 for the out-of-plane, and 5$\times$7 for the in-plane calculations. Note that only the zero bias transmission spectrum is calculated. The Landauer formula is then used to extract the conductance:
\begin{equation}
    G = \frac{2q^2}{h}\int_{-\infty}^{\infty} T(E)\left(-\frac{\partial f(E,E_F,T)}{\partial E}\right) dE
    \label{equ:G}
\end{equation}

%\section*{Author contributions}
%We strongly encourage authors to include author contributions and recommend using \href{https://casrai.org/credit/}{CRediT} for standardised contribution descriptions. Please refer to our general \href{https://www.rsc.org/journals-books-databases/journal-authors-reviewers/author-responsibilities/}{author guidelines} for more information about authorship.

\section*{Conflicts of interest}
There are no conflicts to declare

\section*{Data availability}
The software used for DFT calculation is Quantum ATK \cite{QATK}. The crystal structure used is from the Quantum ATK database. The code for software Atomsk used to create grain boundaries can be found at can be found at \cite{Atomsk}. 
%A data availability statement (DAS) is required to be submitted alongside all articles. Please read our \href{https://www.rsc.org/journals-books-databases/author-and-reviewer-hub/authors-information/prepare-and-format/data-sharing/#dataavailabilitystatements}{full guidance on data availability statements} for more details and examples of suitable statements you can use.

\section*{Acknowledgements}
This work is part of the research project ENERGIZE under grant agreement No. 101194458 funded by the European Union’s Horizon Europe. This work is also supported by the Spanish Government through the projects PID2020-116518GB-I00 funded by MCIN/AEI/10.13039/501100011033 and CNS2023-143727 RECAMBIO funded by MCIN/AEI/10.13039/501100011033, and the European Union NextGeneration EU/PRTR and by Consejería de Universidad, Investigación e Innovación de la Junta de Andalucía, through the P21${\_}$00149 ENERGHENE research project. F.P. acknowledges the funding from the R+D+i project A-ING-253-UGR23 AMBITIONS cofinanced by Consejería de Universidad, Investigación e Innovación and the European Union, under the FEDER Andalucía 2021-2027. Juan J. Palacios would like to thank the Spanish MICINN (grants nos. TED2021-131323B-I00, and PID2022-141712NB-C21), the María de Maeztu Program for Units of Excellence in R\&D (grant no. CEX2018-000805-M), the Comunidad Autónoma de Madrid through the NextGeneration EU from the European Union (MAD2D-CM-UAM7), and the Generalitat Valenciana through the Program Prometeo (2021/017). The views and opinions expressed are those of the authors only and do not necessarily reflect those of the European Union or the European Commission. Neither the European Union nor the European Commission can be held responsible for them.

%%%END OF MAIN TEXT%%%



%%%REFERENCES%%%
\bibliography{main} %You need to replace "rsc" on this line with the name of your .bib file
\bibliographystyle{IEEEtran} %the RSC's .bst file
\end{document}